\documentclass[12pt,preprint]{aastex}
\usepackage{epsf}
\usepackage{lscape}

\usepackage{graphicx}
\usepackage{natbib}
\begin{document}
\title{MEASURING THE MASS DISTRIBUTION IN GALAXY CLUSTERS}

\author {Margaret J. Geller} 
\affil{Smithsonian Astrophysical Observatory,
\\ 60 Garden St., Cambridge, MA 02138}
\email{mgeller@cfa.harvard.edu}
\author{Antonaldo Diaferio}
\affil{Dipartimento di Fisica,
\\Universit\`a degli Studi di Torino, via P. Giuria 1, 10125 Torino, Italy}
\affil{ INFN, Sezione di Torino, via P. Giuria 1, 
\\10125 Torino, Italy}
\email{diaferio@ph.unito.it}
\author{Kenneth J. Rines}
\affil {Department of Physics \& Astronomy, Western Washington University, Bellingham, WA 98225}
\email{kenneth.rines@wwu.edu}
\author{Ana Laura Serra}
\affil{INAF, Osservatorio Astronomico di Torino, via Osservatorio 20, 10025 Pino Torinese (TO), Italy}
\affil{Dipartimento di Fisica, 
\\Universit\`a degli Studi di Torino, via P. Giuria 1, 10125 Torino, Italy}
\affil{ INFN, Sezione di Torino, via P. Giuria 1, 
\\10125 Torino, Italy}
\email {serra@to.infn.it}
\begin{abstract}

Cluster mass profiles are tests of models of structure formation. Only two current observational methods of determining the mass profile, gravitational lensing and the caustic technique,  are independent of  the assumption of dynamical equilibrium. Both techniques enable determination of the extended mass profile at radii beyond the virial radius. For 19 clusters, we compare the mass profile based on the caustic technique with weak lensing measurements taken from the literature. This comparison offers a test of systematic issues in both techniques. Around the virial radius, the two methods of mass estimation agree to within
$\sim 30$\%, consistent with the expected errors in the individual techniques. At small radii, the caustic technique overestimates the mass as expected from numerical simulations. The ratio between the
lensing profile and the caustic mass profile at these radii  suggests that the weak lensing profiles are a good representation of the true mass profile. At radii larger than the virial radius, the lensing mass profile exceeds the caustic mass profile possibly as a result of contamination of the lensing profile by large-scale structures within the lensing kernel. We highlight the case of the closely
neighboring clusters MS0906+11 and A750 to  illustrate the potential seriousness of contamination of the the weak lensing signal by unrelated structures.

\end{abstract}

\keywords{galaxies:clusters individual (MS0906+11, A750) - galaxies:kinematics and dynamics - gravitational lensing - cosmology:observations - cosmology:dark matter} 

\section{Introduction}

Mass profiles of clusters of galaxies are tests of models of
structure formation on scales from 40h$^{-1}$ kpc to 5h$^{-1}$ Mpc. They probe the nature of dark matter
and enable exploration of the link between cosmology and cluster formation.

Navarro, Frenk \& White (NFW: 1997) spurred interest in cluster mass profiles when they demonstrated that the profile for virialized halos has a universal form independent of the initial power spectrum of density fluctuations and of the cosmological parameters. Two parameters, a mass (within a radius
surrounding a region with a specified average density contrast) and a central concentration completely define the NFW profile. NFW predicted greater concentration for lower mass systems reflecting their earlier formation time. More recently, Merritt et al. (2006), Navarro et al. (2010) and others have argued that the shapes of dark matter halo mass profiles are better approximated by the three-parameter Einasto profile (1965; see Coe 2010 for a tutorial comparison of the profiles). The NFW form remains a good approximation; it is thus widely applied to characterize observations. 

Combined strong and weak lensing observations of several massive clusters  suggest a tension between observed cluster mass profiles and theoretical predictions. A set of eight well-studied  clusters
with masses $\gtrsim 5 \times 10^{14}$ h$^{-1}$ M$_\odot$ are more
centrally concentrated than models predict (Broadhurst et al. 2008; Oguri et al. 2009; Sereno et al. 2010; Postman et al. 2012). The observations are impressive because they probe the central region directly and the resulting mass distribution is independent of the detailed dynamical state of the cluster. These results suggest an earlier formation time for massive clusters than predicted by the standard model
(Duffy et al. 2008; Prada et al. 2011). On the other hand,  the presence of strong lensing arcs may select for more centrally concentrated clusters than those typical of the mass range (Hennawi et al. 2007; Oguri \& Blandford 2009; Meneghetti et al. 2010).

Umetsu et al. (2011) combine four clusters with a characteristic mass of $\sim 1.5 \times 10^{15}$h$^{-1}$M$_\odot$ to derive  a single high precision mass
profile covering the range 40h$^{-1}$ kpc to 2.8h$^{-1}$ Mpc. The profile agrees impressively well with the NFW prediction even at radii exceeding the virial radius. The concentration of this summed profile exceeds typical model predictions. Okabe et al. (2010) derive stacked weak lensing profiles for two
larger sets of less massive clusters; the profiles are less concentrated than the Umetsu et al (2011) profile, suggesting that the smaller Umetsu et al. (2011) sample may be biased. Like the Umetsu et al (2011) profile, the Okabe et al (2010) profiles are a a superb match to the NFW form over a large radial range.  

In addition to comparisons with theoretical expectations, weak lensing masses and mass profiles have been tested against other observational approaches. Masses derived from x-ray observations and from 
equilibrium dynamical analyses of cluster redshift surveys  are generally in  impressive agreement with the lensing results (e.g. Irgens et al. 2002; Diaferio et al. 2005; Hoekstra 2007; Okabe et al. 2010). A drawback of these comparisons is that most other observational techniques, unlike lensing, assume dynamical equilibrium and they do not extend to large radius.

Here we compare a set of weak lensing mass profiles from the literature (Hoekstra 2007; Okabe \& Umetsu 2008; Lemze et al. 2008; Umetsu et al. 2009; Okabe et al. 2010; Oguri et al. 2010) with profiles  determined from the caustic technique. Diaferio and Geller (DG97: 1997) and Diaferio (D99: 1999) used  simulations to 
determine  the cluster mass distribution without requiring the equilibrium assumption.  They showed that the amplitude of the trumpet-shaped pattern typical of clusters in redshift space (Kaiser 1987;
Reg\"os \& Geller 1989) is a measure of the escape velocity from the cluster. The mass estimator
based on this identification enables measurement of the cluster mass profile
within the virialized central region and throughout the surrounding infall region. Rines et al. (2012) show that the NFW profile is a good representation of the data on scales $\lesssim 2h^{-1}$ Mpc; on larger scales the profiles appear to steepen.
 
Applications of the caustic technique include the analysis of a sample of nearby clusters
(Rines et al. 2003),  a sample of 72 clusters
included in the Sloan Digital Sky Survey (Rines \& Diaferio 2006; CIRS), a sample of x-ray selected groups (Rines \& Diaferio 2010), and detailed studies of individual systems (e.g. Geller, Diaferio \& Kurtz 1999; Reisenegger et al. 2000; Drinkwater et al. 2001; Biviano \& Girardi 2003; 
Lemze et al. 2009; Lu et al. 2010).
These studies include comparisons with masses determined from the x-ray and from Jeans analysis; in general the masses agree reasonably well. As for the weak lensing profiles,
caustic mass profiles extending beyond the virial radius are 
consistent with the NFW form over a wide range of scales (Geller, Diaferio \& Kurtz 1999;
Rines \& Diaferio 2006; Rines et al. 2012). 
In contrast with the smaller sample of clusters studies with weak lensing, kinematic analysis of these large, complete samples of x-ray selected clusters imply concentrations consistent with theory
(Rines \& Diaferio 2006). These results underscore the importance of sample selection in the determination of mass profiles (see e.g. Postman et al. 2012).

Comparison of kinematic mass profiles with weak lensing offers an
observational test of both techniques. Ultimately, if the systematic biases in these techniques can be understood, this kind of comparison could be a route to testing alternative theories of gravity
(Lam et al. 2012) and the dark matter equation of state (e.g. Faber \& Visser 2006; Serra \& Dominguez Romero 2011).

The number of clusters with weak lensing observations and with the dense spectroscopy necessary for application of the caustic technique has been small.
Lemze et al. (2009)  compare the caustic mass profiles with the weak lensing profiles for  A1689 and find impressive agreement. Umetsu et al. (2010) make a similar detailed comparison for Cl0024+1654 and show that the weak lensing mass significantly exceeds the caustic mass at radii approaching the virial radius. The caustic mass is consistent with earlier weak lensing estimates (Kneib et al. 2003) that measure the mass distribution only in the central halo of this complex system.
Diaferio, Geller \& Rines (2005) used earlier data for three clusters, A2390, MS1358.4+6245, and Cl0024+1654 to show that the profiles are in reasonable agreement. 

The HeCS (Hectospec Cluster Survey; Rines et al. 2012) sample of clusters in the redshift range $0.1 < z < 0.3$ substantially increases the overlap between the set of clusters with extensive spectroscopy and those with weak lensing measurements. Among the 58 clusters in the X-ray flux limited HeCS sample, 
seventeen have published weak lensing mass profiles. Two additional clusters in the CIRS cluster sample drawn from the SDSS also have weak lensing profiles. Here we compare the caustic mass profile estimates for these nineteen systems with the weak lensing estimates.

We review the HeCS and CIRS samples in Section 2. In Section 3  we briefly review the caustic technique we apply. We compare the caustic mass profiles with published weak lensing mass profiles and examine possible sources of disagreement. We discuss the results in Section 4 and conclude in Section 5.

\section {Cluster Surveys and Weak Lensing Mass Profiles}

CIRS (Cluster Infall Regions in SDSS; Rines \& Diaferio 2008) and HeCS (Hectospec Cluster Survey; Rines et al. 2012) are redshift surveys of x-ray flux limited samples of clusters of galaxies.  There are typically more than $\sim 150$ spectroscopically confirmed cluster members inside the turnaround radius. Together these samples include 130 clusters. The CIRS clusters are typically at $z \lesssim 0.1$; the HeCS clusters are in the range $ 0.1 < z < 0.3$. 
Here we briefly review the cluster redshift surveys and sources of the weak lensing profiles we compare with our dynamical estimates.

\subsection {The Cluster Redshift Surveys}

Rines \& Diaferio (2008) extracted the CIRS sample from the Fourth Data Release of the SDSS 
(Adelman-McCarthy et al. 2006).  The cluster sample is x-ray flux limited with $f_x > 3 \times 10^{-12}$ erg s$^{-1}$ cm$^{-2}$ (0.1-2.4 keV). The clusters have redshift $z < 0.1$ and thus the SDSS redshift survey reaches to approximately 
M$^*_{r, 0.1}$ + 1 within each cluster. This photometric limit guarantees a large enough sample of cluster members to determine the cluster boundaries in redshift space. There are 72 clusters in the sample and the clusters contain a total of more than 15,000 members. Approximately a third of the members are projected within the virial radius; the rest are projected outside the virial radius but within the infall region. 

Rines et al. (2012) enlarge the number of spectroscopically well-sampled clusters by using 
the 300-fiber Hectospec on the 6.5-meter MMT to measure redshifts in 58 x-ray selected clusters. These clusters span the redshift range $0.1 < z < 0.3$. The x-ray flux limit is $5 \times 10^{-12}$ erg s$^{-1}$ cm$^{-2}$ in the ROSAT band (Ebeling et al. 1998; B\"ohringer et al. 2000) . 

To maximize the efficiency of the Hectospec observations, targets for spectroscopic observations are within $\pm 0.3$ magnitudes of the red sequence at the
cluster redshift. This strategy typically yielded $\sim 150-200$ members in each cluster with two pointings of the 300-fiber Hectospec (Fabricant et al. 2005) per cluster. The HeCS survey includes more than 20,000 new redshifts; among these more than 10,000 are cluster members. 

For all 130 clusters in the CIRS and HeCS samples, the redshift survey is 
dense enough to define the boundaries of the cluster in redshift space. We have applied the
caustic technique uniformly to each set of clusters. Here we focus on the subset of nineteen clusters in the two surveys that have mass profiles derived from weak lensing observations. 

\subsection {Weak Lensing} 
\label{weaklensing}

We searched the literature for weak lensing mass profiles for clusters in both the CIRS and HeCS samples. We found nineteen matches; 2 in CIRS and 17 in HeCS. Table 1 lists the 
clusters and the source of the weak lensing profile.

Two clusters in the CIRS sample have weak lensing mass profiles derived from Subaru data
(Okabe \& Umetsu 2008; Umetsu et al. 2009). Okabe \& Umetsu  (2008) derived mass profiles for seven merging clusters; three of these systems are in the HeCS survey (see Table 1) and one (A1750) is in the CIRS sample. Umetsu et al. (2009) also studied the complex cooling front system A2142 (in CIRS). The x-ray properties of these clusters suggest that their central regions are not in dynamical equilibrium. 

Among the 17 clusters in HeCS with weak lensing mass profiles, 15 are also derived from
Subaru observations (Okabe \& Umetsu 2008; Okabe et al. 2010; Umetsu et al. 2009). Six clusters have weak lensing profiles derived from CFHT observations (Hoekstra 2007); among these, three are clusters that also have Subaru
weak lensing profiles. We use the differences in the
Subaru- and CFHT-derived profiles as a measure of the uncertainty in the lensing profile. For one cluster, A1689, there is a CFHT-derived profile (Hoekstra 2007) and a profile derived from HST data (Lemze et al. 2008).  The clusters A1763 and MS0906+11 have  profiles derived from CFHT data alone (Hoekstra 2007). In cases where there is more than one profile derived from Subaru observations, we chose the most recent analysis. Table
\ref{tbl:weaklens} summarizes the sources of the weak lensing profiles. 

\begin{figure}[htb]
\centerline{\includegraphics[width=7.0in]{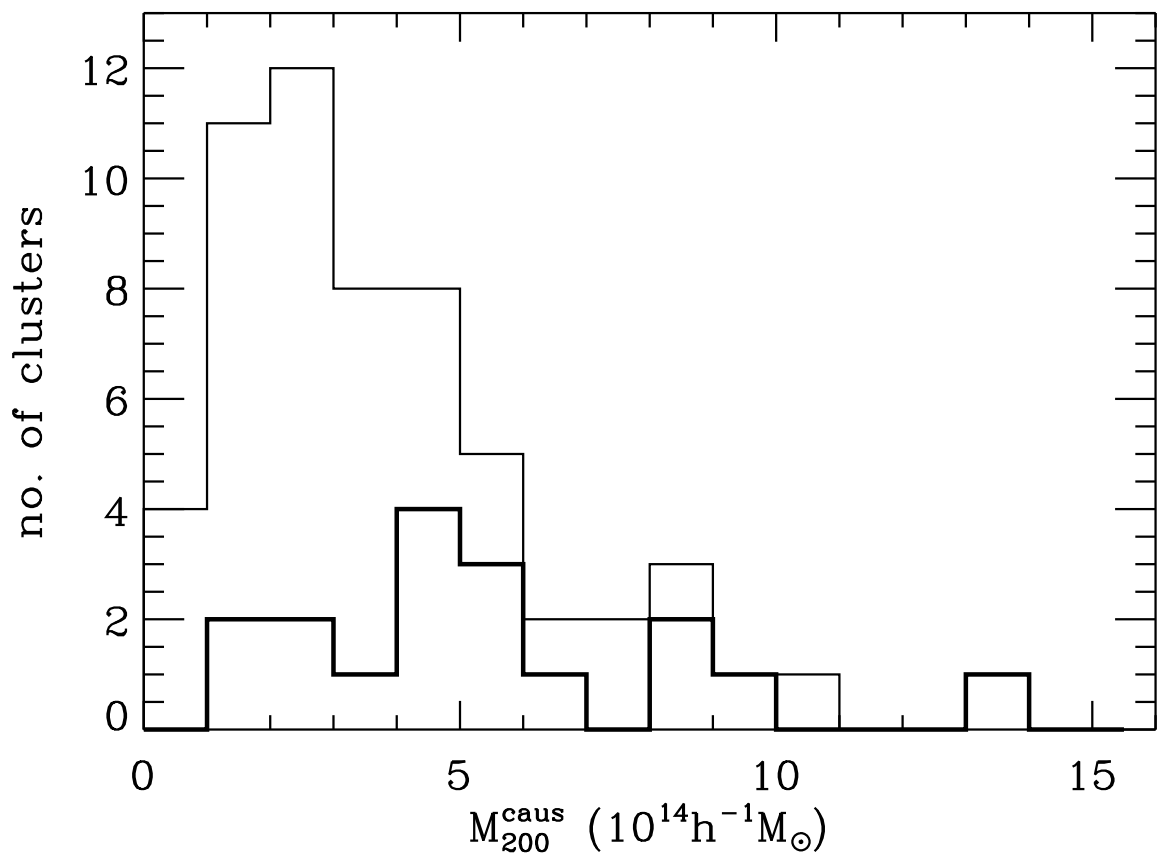}}
\vskip -2ex
\caption{Distribution of caustic masses, M$_{200}^{caus}$ for the HeCS sample (thin line) and for the
subsample with weak lensing mass profiles (heavy line). The observed lensing clusters tend to be more massive. 
\label{fig:mass-hist.ps}}
\end{figure}

Figure \ref {fig:mass-hist.ps}
shows the distribution of masses for HeCS clusters (determined from the caustic technique; Rines et al. 2012) within the radius containing an average density 200 times the
critical density, M$_{200}^{caus}$ (thin histogram). The heavy histogram shows the distribution for the subset with weak lensing profiles in the literature. Clearly the systems with weak lensing measurements tend to be more massive. This shift toward more massive clusters in the lensing sample results from selection. The HeCS clusters are a flux limited sample; the clusters with weak lensing profiles are generally among the most intrinsically luminous (and thus generally most massive) x-ray clusters.

For the sample of seventeen HeCS clusters, the fifteen Subaru profiles were analyzed by collaborating groups of investigators who use consistent techniques; these profiles thus provide a reasonably uniform testbed for the caustic profiles we derive.  

\begin{figure}[htb]
\centerline{\includegraphics[width=7.0in]{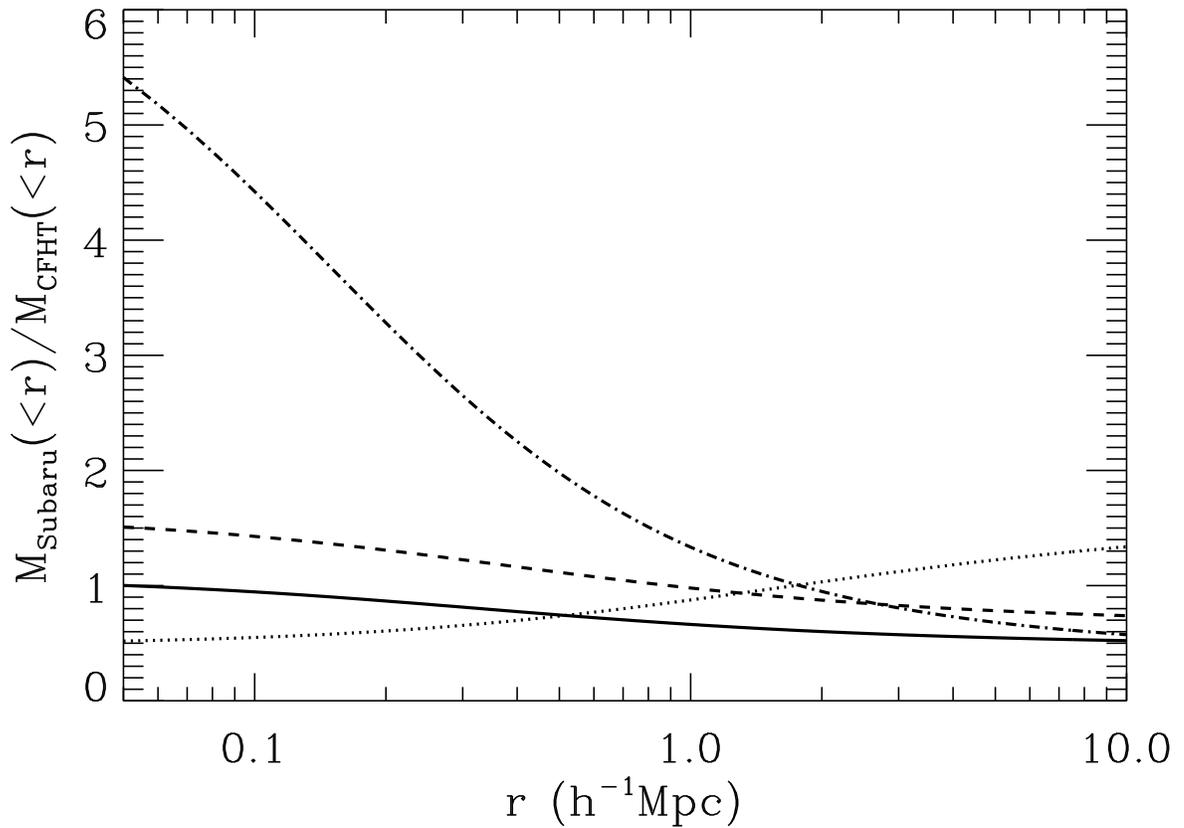}}
\vskip -2ex
\caption{Ratio of weak lensing profiles for clusters observed with more than one facility.
We plot M$_{Subaru}$/M$_{CFHT}$ for all cases except A1689
where we plot M$_{HST}$/M$_{Subaru}$. We plot profile ratios for A1689 (dot-dashed curve), A2219 
(dashed), A267 (solid), and A963 (dotted). These ratios suggest that systematic errors in the lensing profiles are $\sim 50 \%$.
\label{fig:lens-ratio.ps}}
\end{figure}

Figure \ref {fig:lens-ratio.ps} shows the ratio between the two weak lensing profiles for each of the four clusters observed with Subaru and CFHT or with CFHT and HST. The Figure shows that for ground-based facilities the profiles
differ by $\lesssim 50\%$ throughout  the radial range. The profile derived from HST data is much more centrally concentrated than the ground-based profile for A1689.

Systematic errors in weak lensing mass profiles may originate from both astrophysical and data reduction issues. We cannot say which, if any, of these issues affect the
relative profiles in Figure \ref {fig:lens-ratio.ps}, but we review the possibilities. The three issues that appear to produce the largest systematic effects are astrophysical (Okabe et al. 2010;
Oguri et al. 2010): contamination of the source population by 
faint galaxies in the cluster, errors in the source redshift, and projection of foreground/background structures.

Dilution of the source catalog by faint cluster members is a potential systematic with an obviously decreasing impact as a function of distance from the cluster center. At small radii where the
contamination is potentially largest, the dilution leads to an underestimate of the concentration of the mass profile. The virial mass estimate, primarily sensitive to distortions at large radius relative to the cluster center, is relatively unbiased by this effect.

The source redshift controls the overall amplitude of the distortion signal. In general, the claimed error in the average distance ratio is only 5 to 10\%. 

Because weak lensing measures the total mass within the lensing kernel and projected within the aperture, foreground and/or background structures along the line-of-sight affect the measurement
(e.g. Hoekstra 2001; Hoekstra 2003; White et al. 2002; de Putter \& White 2005; Hoekstra et al. 2011). Umetsu et al (2011) show (their Figure 1) that the cosmic noise is an increasing fraction of the lensing signal at
larger distance from the cluster center. Oguri et al. (2010) and Okabe et al. (2010) argue that any biases introduced by projected large-scale structure are generally insignificant. In thier study of A2261,  Coe et al. (2012) remove superposed background structures; this procedure reduces the virial mass by only $\sim 7$\% and increases the concentration by only $\sim 5$\%. 
However, the effect can be large in some cases; 
we demonstrate a textbook case in Section \ref{ms0906}.

Another issue is departure from spherical symmetry. Most masses derived from weak lensing data are based on azimuthally averaged (1D) profiles. Oguri et al. (2010) and Okabe et al. (2010)  make full use of the 2-dimensional weak lensing data and compare the shear pattern with predictions of elliptical models. Oguri et al. (2010; Figure 7) show that the 
1D and 2D virial masses are remarkably consistent with one another: they differ by only 10\% in the mean and the scatter is $\lesssim 20\%.$ Although
Oguri et al. (2010) provide 2D weak lensing profiles for some of the HeCS clusters, we use the 1D profiles because they are available for  18/19 clusters and because the caustic technique also assumes spherical symmetry. For the cluster Abell 689 where no 1D profile is available, we use the 2D profile from Oguri et al. (2010).   

Less important issues affecting the weak lensing profiles include differences in the identification of the cluster center and errors
in measurement of the shapes of the sources. Centering errors are generally small and various methods of measuring the shapes of the sources differ by $\sim 10\%$.

Regardless of the source of the differences, the difference between lensing profiles for a single cluster are comparable to the difference between the caustic and lensing mass estimates (see Section 
\ref {caustics-lensing}). We discuss the fundamental limits for the caustic technique in Section \ref{caustics}.

\section {Cluster Mass Profiles: Caustics }
\label {caustics}
For comparison with the weak lensing profiles, we use the caustic technique (DG97; D99) to derive mass  profiles from dense kinematic data for each of the clusters. 

Like weak lensing, the caustic technique makes no assumption about the dynamical equilibrium of the system. In contrast with weak lensing, the caustic technique measures the three-dimensional distribution of mass rather than the projected mass. The caustic technique also extends to large radius and provides a test of weak lensing results outside the virial radius. Structure along the line-of sight is generally resolved by the redshift survey; the caustic technique is thus insensitive to it.  Like the most straightforward applications of weak lensing, the caustic technique assumes spherical symmetry. More sophisticated applications of weak lensing mass estimates do not necessarily assume azimuthal symmetry; they provide an estimate of the projected surface mass density as a function of position on the sky (see Okabe et al. 2010; Oguri et al. 2010).

In redshift space, a cluster of galaxies appears as a trumpet-shaped pattern (Kaiser 1987; Reg\"os \& Geller 19889). DG97 and D99 
demonstrated that for clusters forming hierarchically,  the boundaries of this sharply defined pattern (termed caustics) in redshift space (a projection of phase space) can be identified with the escape velocity from the cluster. This identification provides a route to estimation of the cluster mass profile assuming spherical symmetry. 

The amplitude of the caustics $A(r)$ is half the distance between the boundaries of the cluster in redshift space. With the assumption of spherical symmetry the gravitational potential $\phi(r)$ and the
caustic amplitude $A(r)$ are related by

$$ A^2(r) = -2\phi(r){{1 - \beta(r)} \over {3 -2\beta(r)}}$$

\noindent where $\beta(r)$ is the anisotropy parameter, $\beta(r) =  1 - \sigma_\theta^2(r)/\sigma_r^2(r)$ where $\sigma_\theta$ and $\sigma_r$ are, respectively, the tangential and radial velocity dispersions.

DG97 show that the mass of a spherical shell within the infall region is the integral of the square of the caustic amplitude $A(r)$:

$$GM(<r) - GM(<r_0) = F_\beta \int_r^{r_0} A^2(x)dx$$

\noindent where $F_\beta \simeq 0.5$ is a filling factor with a value estimated from numerical simulations. We approximate $F_\beta$ as a constant; variations in $F_\beta$ with radius lead to some systematic uncertainty in the mass profile we derive from the caustic technique. We include these
issues in our assessment of the intrinsic uncertainties and biases in the technique
(Serra et al. 2011).

The first step in applying the caustic technique is identification of the cluster center. We isolate the cluster initially by selecting all of the galaxies in our redshift survey that lie within 
10h$^{-1}$ Mpc and 5000 km s$^{-1}$ of the nominal x-ray cluster center. We then construct a binary
tree based on pairwise estimated binding energies. We use the tree to identify the largest cluster in the field and we adaptively smooth the distribution of galaxies within this cluster to identify its center (see D99 and Serra et al. 2011 for detailed descriptions of this process).

\begin{figure}[htb]
\centerline{\includegraphics[width=7.0in]{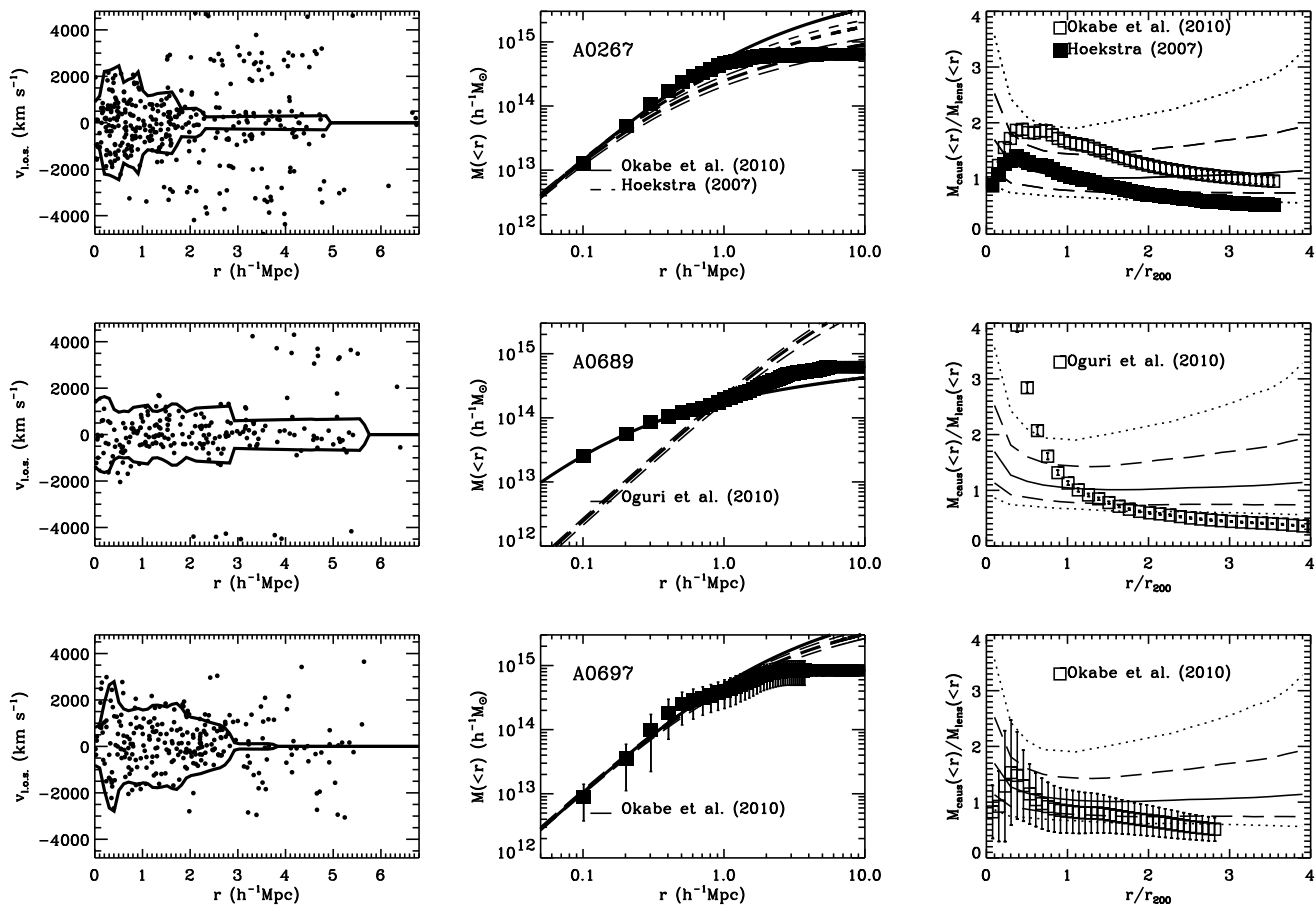}}
\vskip -2ex
\caption{{\it Left column}: Caustic diagrams for individual clusters in the HeCS sample. The vertical axis is the rest frame line-of-sight velocity relative to the hierarchical cluster center; the horizontal axis is the projected distance from the hierarchical center. Each point represents a galaxy with a redshift in the HeCS sample, the
curves are the caustics. {\it Center column:} Mass profiles for individual clusters. Points with error bars are the caustic estimate, the solid curve is an NFW fit to the caustic profile within 1~$h^{-1}$~Mpc, and dashed curves are the NFW fits (heavy) and
1$\sigma$ errors (light) for the corresponding weak lensing profiles with the source indicated in the panel. {\it Right column}: Ratio of caustic mass profiles and weak lensing mass profiles for individual clusters. Points with error bars give the ratio. For cases with more than one weak lensing profile, the symbols 
give the source as indicated in the panel. The solid curve is the ratio between the caustic profile
and true mass profiles computed from N-body simulations (Serra et al. 2011). The dashed and dotted curves are the 1$\sigma$ and 2$\sigma$ limits on the ratio derived in the N-body simulations.
\label{fig:caustics1.ps}}
\end{figure}

\begin{figure}[htb]
\centerline{\includegraphics[width=7.0in]{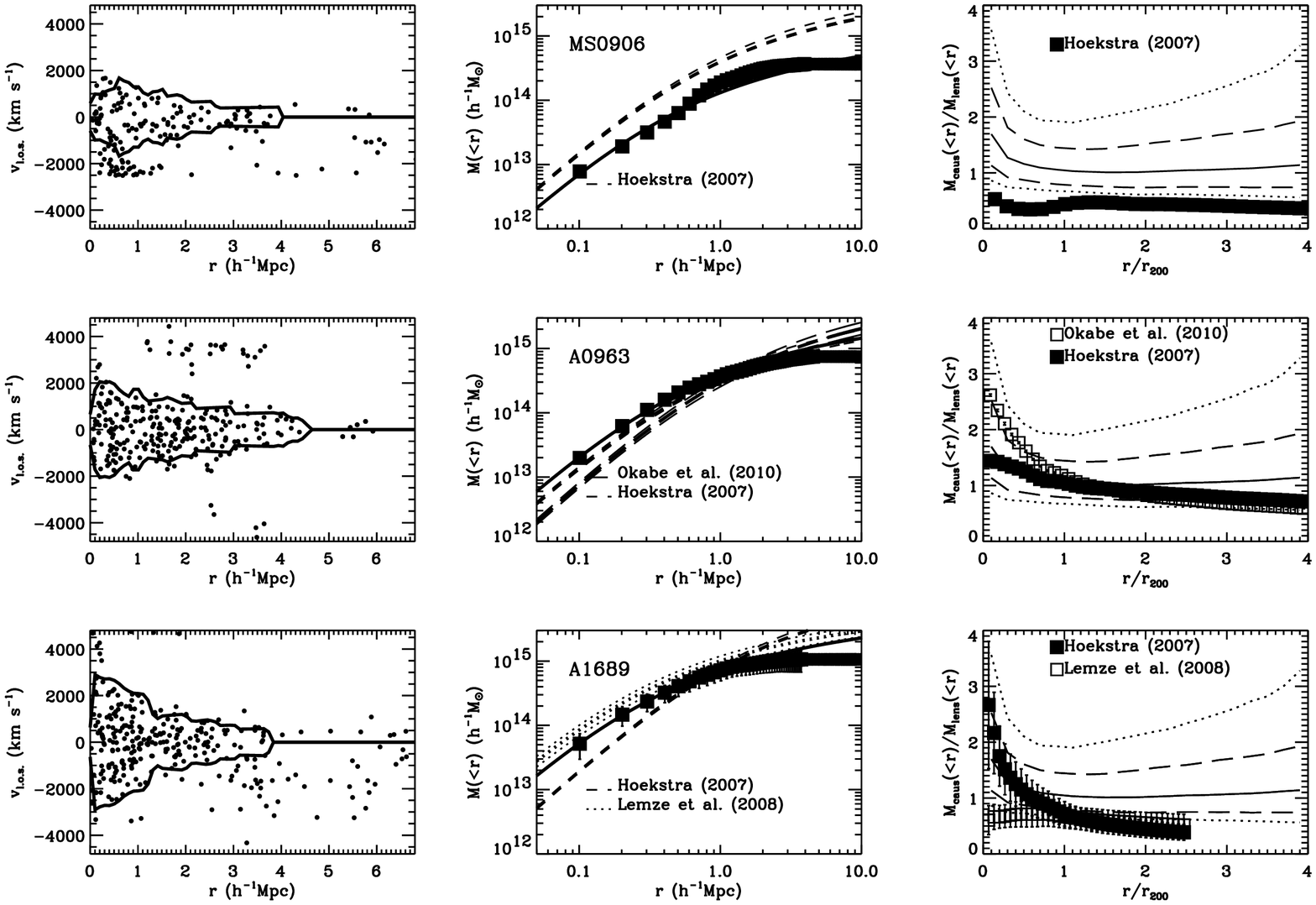}}
\vskip -2ex
\caption{Caustic diagrams (right column), mass profiles (center column), and caustic to lensing profile ratios
(left column) for individual clusters in the HeCS sample as in Figure 3. 
\label{fig:caustics2.ps}}
\end{figure}

\begin{figure}[htb]
\centerline{\includegraphics[width=7.0in]{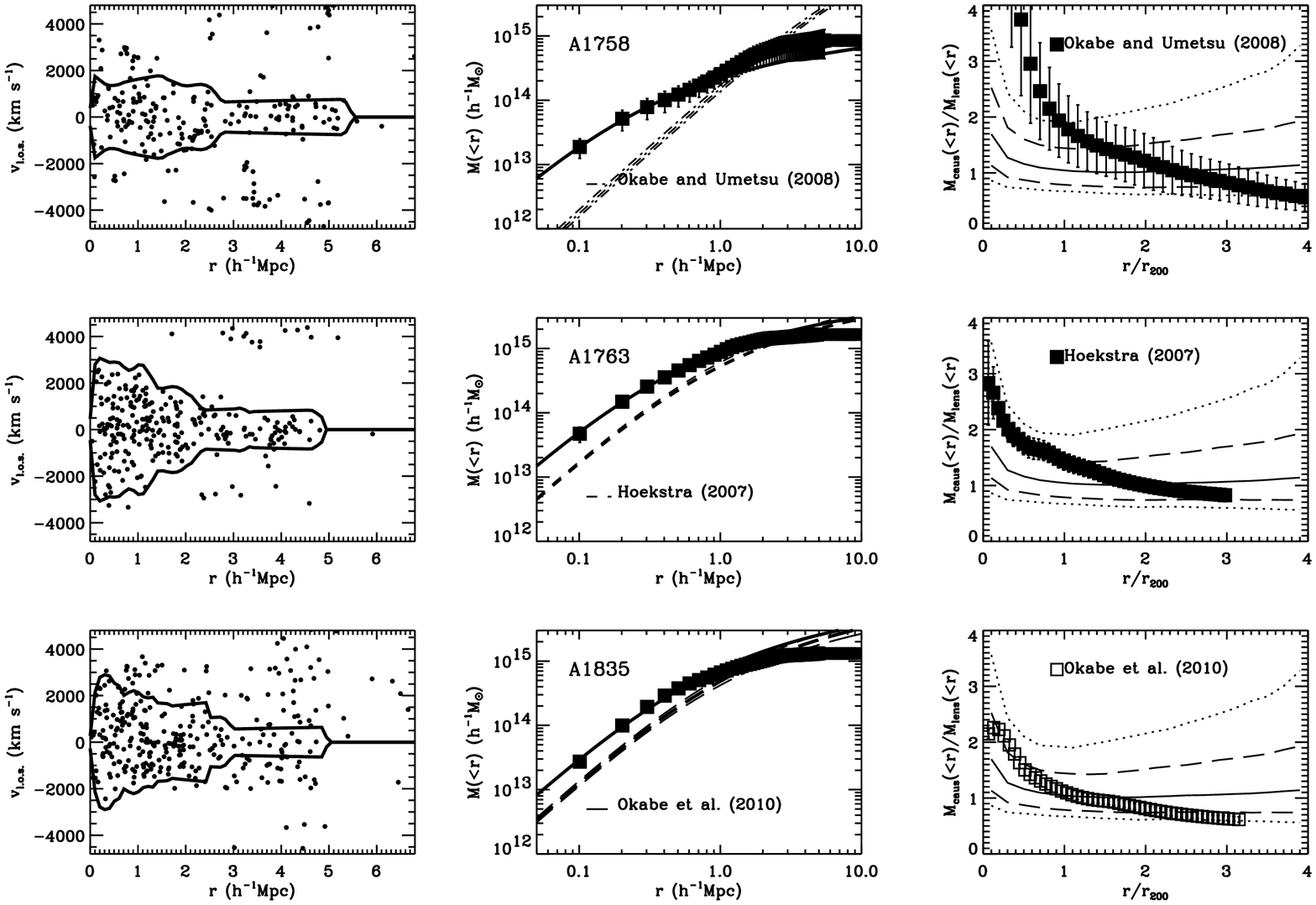}}
\vskip -2ex
\caption{Caustic diagrams (right column), mass profiles (center column), and caustic to lensing profile ratios
(left column) for individual clusters in the HeCS  sample as in Figure 3. 
\label{fig:caustics3.ps}}
\end{figure}

\begin{figure}[htb]
\centerline{\includegraphics[width=7.0in]{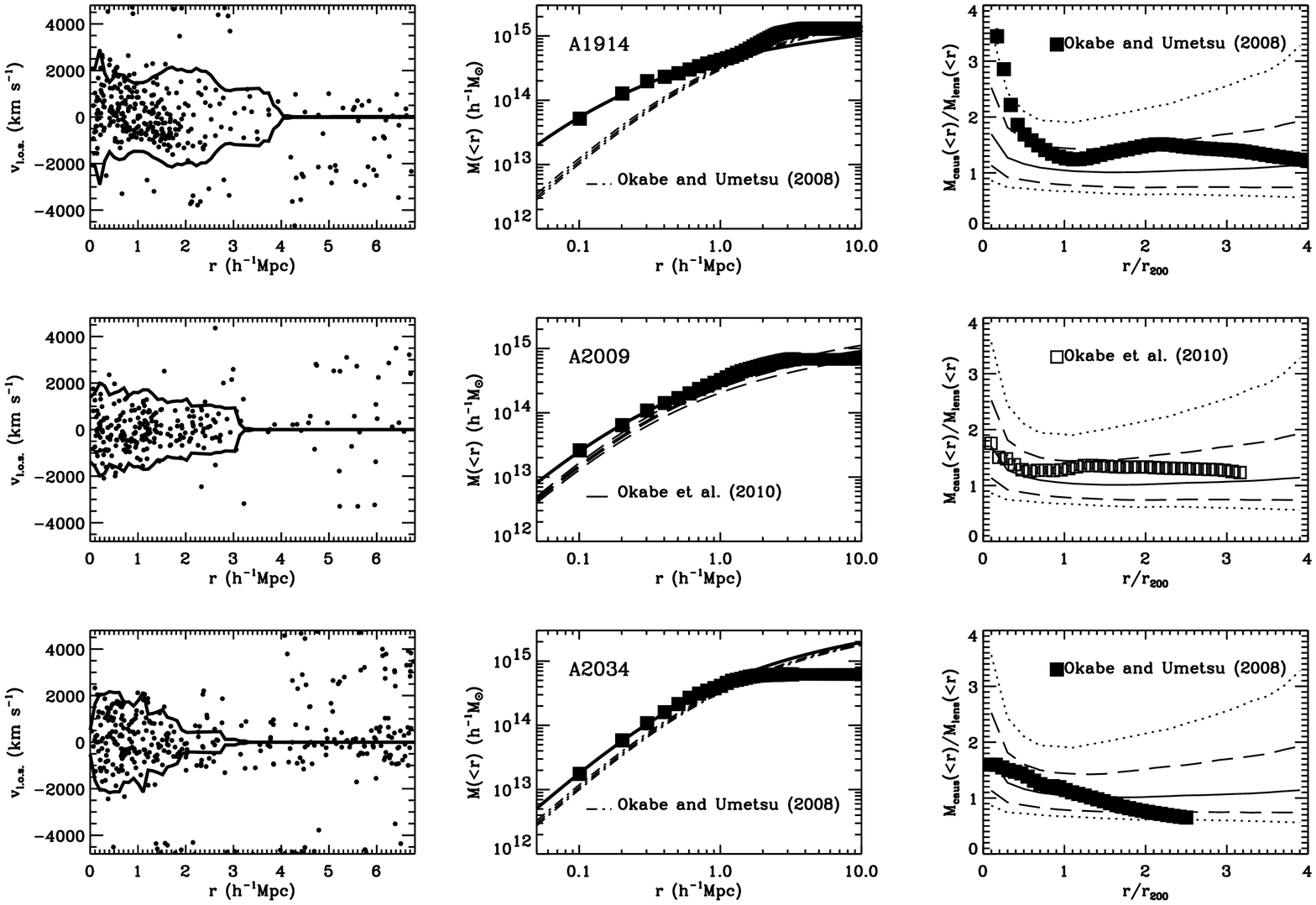}}
\vskip -2ex
\caption{Caustic diagrams (right column), mass profiles (center column), and caustic to lensing profile ratios
(left column) for individual clusters in the HeCS sample as in Figure 3. 
\label{fig:caustics4.ps}}
\end{figure}

\begin{figure}[htb]
\centerline{\includegraphics[width=7.0in]{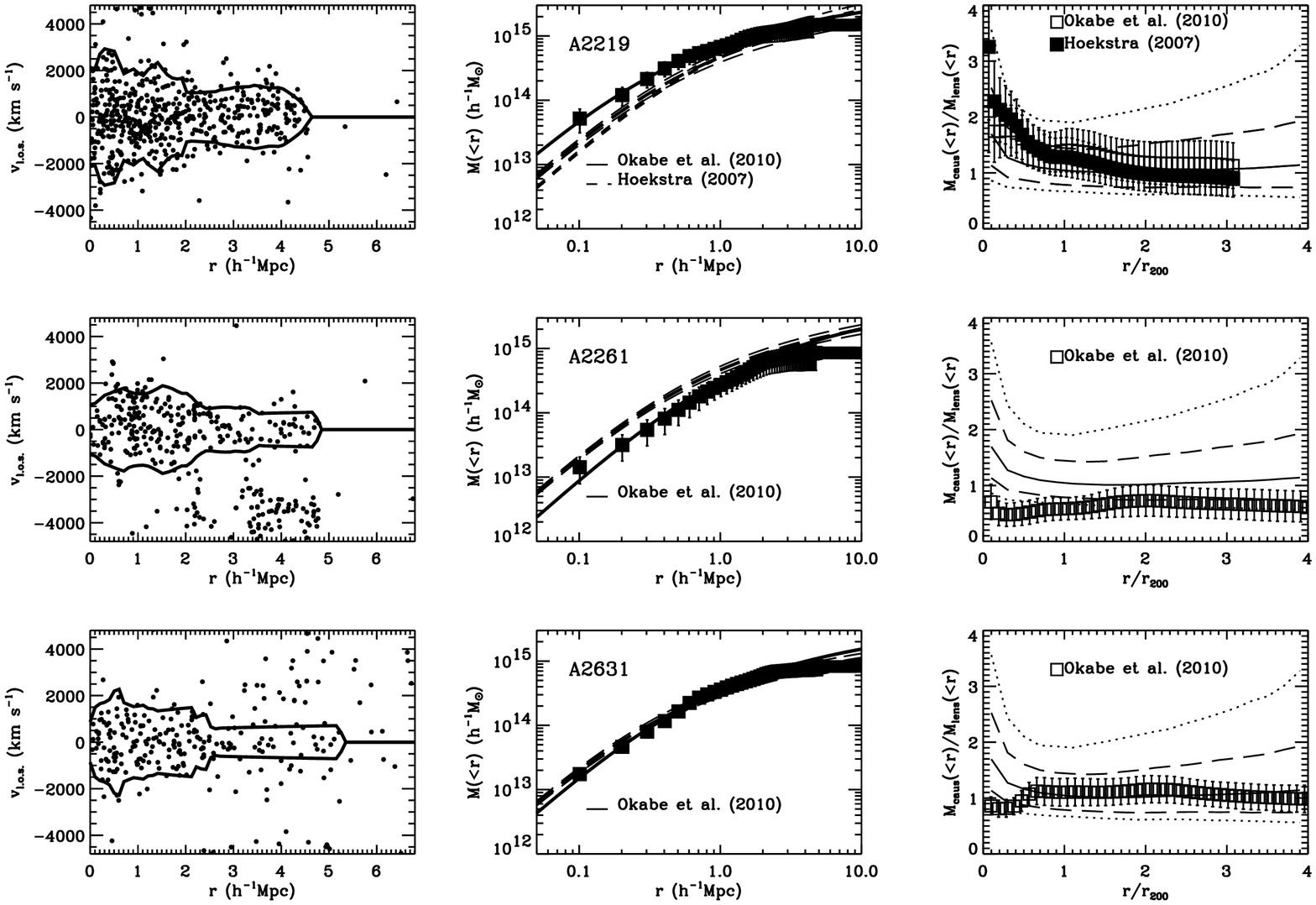}}
\vskip -2ex
\caption{Caustic diagrams (right column), mass profiles (center column), and caustic to lensing profile ratios
(left column) for individual clusters in the HeCS sample as in Figure 3. 
\label{fig:caustics5.ps}}
\end{figure}

\begin{figure}[htb]
\centerline{\includegraphics[width=7.0in]{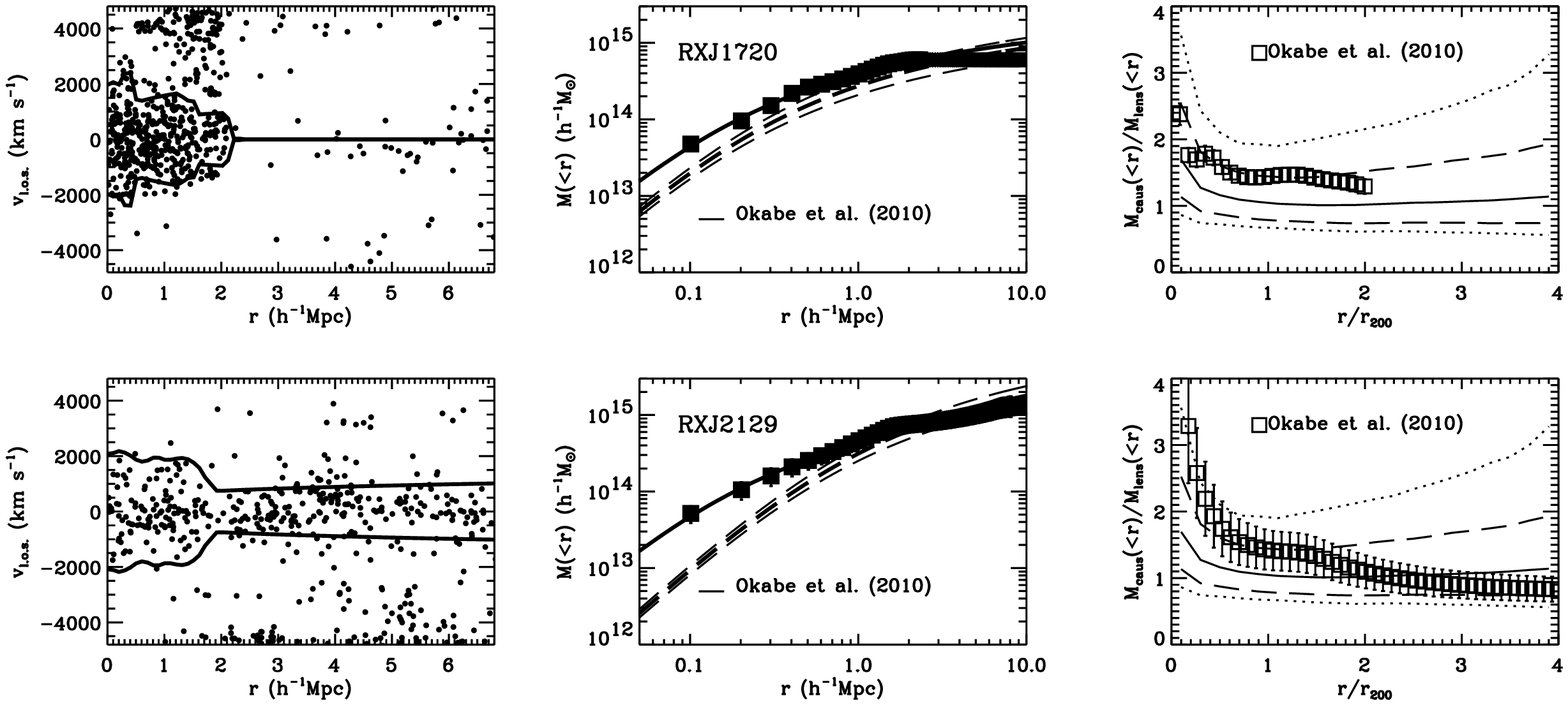}}
\vskip -2ex
\caption{Caustic diagrams (right column), mass profiles (center column), and caustic to lensing profile ratios
(left column) for individual clusters in the HeCS sample as in Figure 3. 
\label{fig:caustics6.ps}}
\end{figure}

\begin{figure}[htb]
\centerline{\includegraphics[width=7.0in]{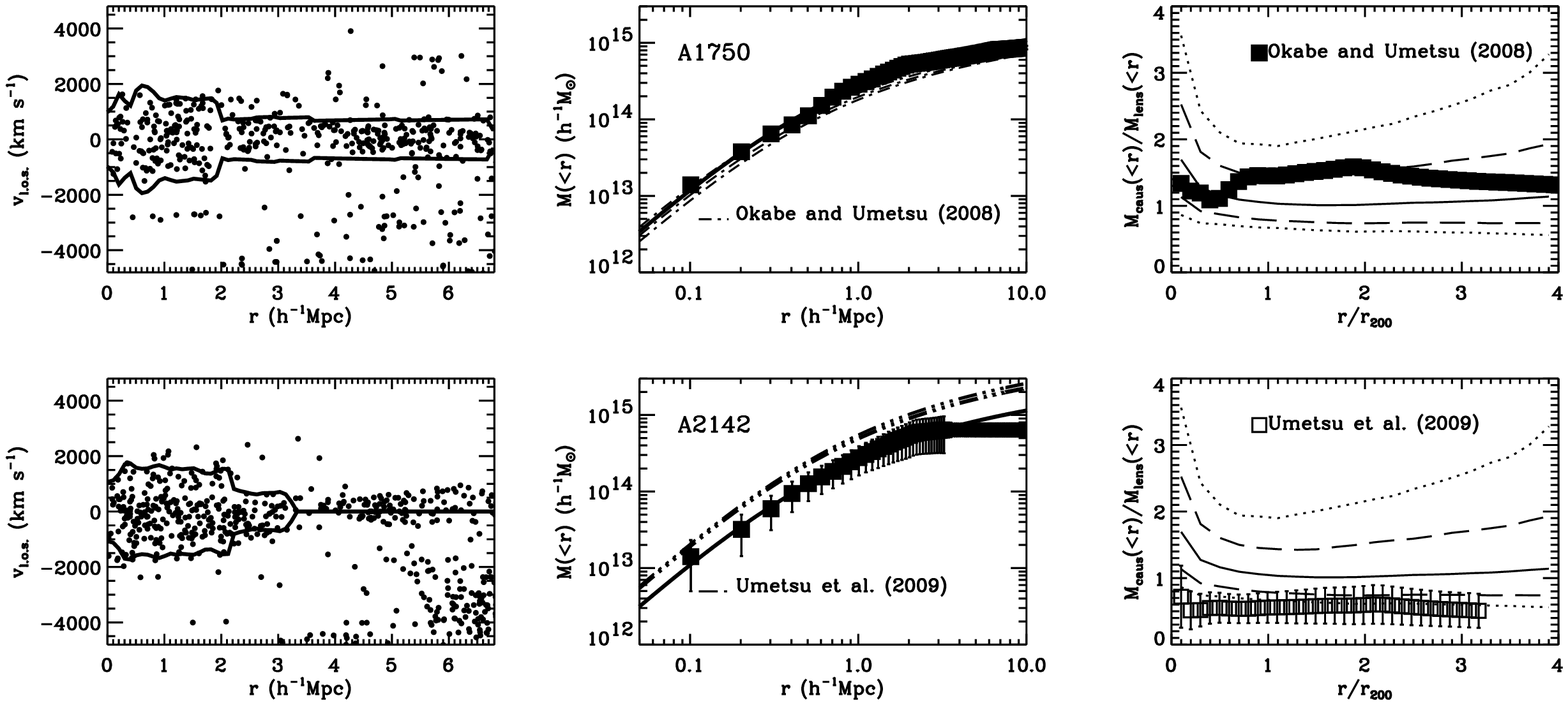}}
\vskip -2ex
\caption{Caustic diagrams (right column), mass profiles (center column), and caustic to lensing profile ratios
(left column) for individual clusters in the CIRS sample as in Figure 3. 
\label{fig:caustics7.ps}}
\end{figure}

Once we have identified a center we can plot the distribution of galaxies in azimuthally summed
phase space. This effective azimuthal averaging smooths over small-scale substructure particularly at large projected radius. Figure \ref {fig:caustics1.ps} --- Figure \ref{fig:caustics7.ps} show the distribution of galaxies in the rest frame line-of-sight velocity versus projected spatial separation plane for the nineteen clusters in this study. The expected trumpet-shape pattern centered on the mean cluster velocity is evident in all cases. 

To measure the amplitude $A(r)$ of the phase space signature of the cluster, we smooth the patterns 
in Figure \ref{fig:caustics1.ps}---\ref{fig:caustics7.ps} and identify a threshold in phase-space density as the edge of the caustic envelope. 
We define the threshold $\kappa$ by solving the equation 
$|\langle{v_{esc}}^2\rangle_{\kappa,R} -\langle{v^2}\rangle_R|=0$ 
where $R$ is a virial-like radius and  $v_{esc}$
is the escape velocity at radius $R$ (D99
and Serra et al. 2011).

In a real cluster, the values of the upper, $A^+(r)$,  and lower, $A^-(r)$, caustic amplitude in the redshift diagram are not identical. Because the caustics of a spherical system are identical, we adopt the smaller value of the two values $A^+(r)$ and $A^-(r)$ as our estimate of $A(r)$.

To compute the shape of the caustics, according to the algorithm described in D99, we choose a smoothing parameter $q = 25$ where $q$ is the scaling between the velocity and radial smoothing in the adaptive kernel estimate of the underlying phase space distribution. For example, a particle with a smoothing window of 0.04$h^{-1}$ Mpc in the spatial direction has a 100 km s$^{-1}$ smoothing window along the velocity direction.
Variations of a factor of 2 in $q$ have essentially no effect on the results (Geller at al. 1999;
Rines et al. 2000, 2002; Rines \& Diaferio 2006). The solid lines in Figures 
\ref{fig:caustics1.ps} --- \ref{fig:caustics7.ps} show the caustics we compute with this procedure.

D99 and Serra et al (2011) investigate the caustic method in detail; they evaluate the uncertainties and systematic biases by applying  the technique to clusters in N-body simulations. On average the caustic recovers the mass profile without any systematic bias and a 1$\sigma$ error of
about $\sim 50$\% in the range $\sim(0.6-4)r_{200}$ ($r_{200}$ is the radius that encloses a mean density 200 times the critical density).  There is a bias toward overestimating the mass at radius smaller
than $\sim 0.6r_{200}$ by $\sim 70$\% at most;
this bias results primarily from the assumption of a constant F$_\beta$. Projection effects, a limitation on every mass estimation technique at some level, are the main source of scatter in the caustic mass estimates.

Table \ref{tbl:causticmass} summarizes the results of the application of the caustic technique to the
nineteen CIRS/HeCS clusters. The caustic mass profiles allow direct estimation of $r_{200}^{caus}$ and the mass within this radius, $M_{200}^{caus}$. Table \ref{tbl:causticmass} lists $r_{200}^{caus}$ and $M_{200}^{caus}$. The Table also lists $N_{caus}$, the number of galaxies within the caustics. These clusters are all sampled well enough for robust application of the technique (see Serra et al. 2011).

We also fit the caustic mass  profiles to the analytic NFW profile within 1~$h^{-1}$~Mpc 
from the cluster center, a typical radius delimiting
the region where the NFW profile is expected to hold:

$$ M(<r) = {M(a)\over{ln(2)-1/2}}{\left[ {ln(1 + {r\over{a}})}-{r\over{a + r}}\right]}$$

\noindent where $a$ is the scale radius and $M(a)$ is the mass with the scale radius. We fit
$M(a)$ rather than the characteristic density $\delta_c$ ($M(a) = 4\pi\delta_c\rho_c{a^3}[ln(2)-1/2]$
where $\rho_c$ is the critical density) because $M(a)$ and $a$ are much less correlated than
$\delta_c$ and $a$ (Mahdavi et al. 1999). Table \ref{tbl:causticmass} lists $M_{200}^{NFW}$,
$r_{200}^{NFW}$ and the NFW concentration parameter $c_{200}$.

\section {Mass Profiles: Caustics and Weak Lensing}
\label{caustics-lensing}
Weak lensing and the caustic technique are fundamentally different measures of the mass distribution in a cluster. Weak lensing measures the total projected mass density within the lensing kernel;
the caustic technique measures the mass within a given radius modulo the effects of both geometric and velocity anisotropy. Both techniques enable the measurement of a mass profile to large projected radius and both are independent of the assumption of dynamical equilibrium.

Comparison of observed mass profiles derived from these two techniques may elucidate the uncertainties and systematic problems in both approaches. Independent structures along the line-of-sight and within the lensing kernel bias the weak lensing mass profiles (e.g. Hoekstra 2003; Coe et al. 2012). The bias is increasingly serious at
large radius where the area on the sky is larger and the projected mass density within the cluster is lower. Application of the caustic mass estimation technique is limited by departures from spherical symmetry (extension along the line-of-sight is also a problem for weak lensing mass estimates)
and by lack of knowledge of the velocity anisotropy. The assumption of a constant $F_\beta$ (which includes the velocity anisotropy information) leads  to an overestimate of the mass at small radii
within $\sim 0.6r_{200}$ (D99; Serra et al. 2011). 

Here we compare caustic and weak lensing profiles with an eye toward exploring the relative systematic 
issues in the lensing and kinematic mass estimates. Figures \ref{fig:caustics1.ps} --- \ref{fig:caustics6.ps} show redshift space diagrams (left column) for the
seventeen HeCS clusters ordered in right ascension. Figure \ref{fig:caustics7.ps} shows the 
redshift space diagrams for the two CIRS clusters. 
The solid lines in these diagrams locate the caustics according to the prescription of D99. The central panel shows the caustic mass profile (points with error bars), the NFW fit to the
caustic profile (solid line) and the NFW fits quoted for the lensing profiles from the literature (dashed and dash-dotted lines). In most cases the caustic mass profile lies above the lensing mass profile at small radius and below at large radius. The profiles generally cross around the virial radius where the two techniques should yield the same result.  In some cases (particularly A2631) the caustic mass profile tracks the weak lensing profile. In two cases, the caustic mass profile lies below the lensing mass profile throughout the range (MS0906 +11 and A2142). There is no obvious reason for the discrepancy in A2142; we discuss the remarkable case of MS0906 +11 in Section \ref{ms0906}.

The data points with error bars in the right panel for each cluster show the ratio between the caustic mass profile, $M_{caus}(<r)$,
and the weak lensing profile, $M_{lens}(<r)$ as a function of projected radius, $r/r_{200}$. 
For the caustic mass profile, we use $r_{200}$ derived from the NFW fit.
The solid, dashed and dotted curves are the median ratio between the caustic profile
and the true profile of a sample of synthetic clusters extracted from an N-body simulation 
(solid curve), along with the 1$\sigma$
(dashed) and 2$\sigma$ (dotted) error range (Serra et al. 2011). These curves demonstrate the results described above in Section \ref{caustics}. This comparison shows that the caustic profile overestimate in the central region is a systematic issue in the technique. However, the underestimate at large radius relative to weak lensing does not reflect an inherent bias in the caustic technique. This latter issue is most probably a bias in the weak lensing profiles at large radius where superposition of large-scale structure is a more serious issue than at small radius simply as a result of the large volume included within the lensing kernel at large radius. 

For the A267, A1689, and A963 we show the ratio between the caustic mass and the weak lensing mass for two different observations (Figure \ref{fig:lens-ratio.ps}). These plots underscore the conclusion from Figure \ref{fig:lens-ratio.ps}. At small radius, the differences between weak lensing profiles derived by different observers on different facilities can be comparable with  the difference between the caustic mass profile and the weak lensing profile. However, the differences between weak lensing profiles do not appear to be systematic from this admittedly very small subsample of four systems.

\subsection {Comments on Individual Clusters}

All of the clusters we consider are selected from x-ray cluster catalogs. HeCS and CIRS are
x-ray flux-limited samples of clusters. We comment here on  the known ``irregularities'' of 8 of the clusters in our weak lensing comparison sample. We reserve discussion of a ninth system, MS0906+11, for Section \ref{ms0906}. 

{\it A689:} This cluster is actually below the flux limit of the HeCS sample. The original
inclusion of A689 in the ROSAT Brightest Cluster Sample (Ebeling et al. 1998) resulted from a superposed BL Lac later
pinpointed by a Chandra observation (Giles et al. 2012). The low mass of this cluster is consistent with the lower x-ray luminosity (Rines et al. 2012). This issue should not have any effect on the comparison of caustic and weak lensing mass profiles.

Okabe et al. (2010) do not fit a mass profile to their weak lensing data for this cluster because the mass map shows prominent substructures.  
Oguri et al. (2010) provide the fit we use for comparison with the caustic mass profile, but caution that the NFW model fit is unacceptable. Their value of $c_{vir} = 0.41$ does not make physical sense
(as they indicate).  

These issues in interpreting the weak lensing data may account for the very large difference between the caustic and weak lensing profiles within $\sim r_{200}$ (Figure \ref{fig:caustics1.ps}); the caustics, in contrast, are stable and well-defined. The caustic mass profile is insensitive to the substructure. The ratio between the caustic and weak lensing profiles  is among the few that are well outside the ratio between the true and caustic mass profiles derived from N-body calibrations.

{\it A1750:} A1750 is binary x-ray cluster (e.g. Forman et al. 1981; Belsole et al. 2004). We compare
our caustic mass estimate with Okabe \& Umetsu's (2008) lensing profile for the most luminous x-ray component, A1750C.
The hierarchical center for the caustic mass calculation is very close to this x-ray peak in both position and velocity.

{\it A1758:} A1758 is a merger of multiple x-ray clusters. B\"ohringer et al (2000) and others separate the cluster into A1758N and A1758S. The lensing analysis of Okabe \& Umetsu (2008) shows that A1758N consists of two separate clusters A1758N:C and A1758N:SE. They center their mass profile on
A1758N:C (labeling it as A1758N). Because A1758 is one of the most distant clusters in HeCS, it is not very densely sampled. Nonetheless, the caustics are visible and the hierarchical center is nearly coincident with the brightest galaxy in A1758N:C. This center is consistent with the center for the weak lensing analysis.

The weak lensing map of Okabe \& Umetsu shows prominent substructures. As in the case of A689, these substructures my be responsible, at least in part, for the very large difference between tha cautic and weak lensing mass profiles within $r_{200}$.

{\it A1914:} X-ray observations suggest that A1914 is a major merger in progress (Govoni et al. 2004). Although the caustic pattern is visible in Figure \ref{fig:caustics4.ps}, the
distribution of galaxies within the region is odd. In spite of this odd distribution, the caustic and weak lensing agree to within the expected 1$\sigma$ scatter for the caustic mass profiles around r$_{200}$.

{\it A2034:} A2034 (Kempner, Sarazin \& Markevitch 2003) is a cold front cluster. The mass map (Okabe \& Umetsu 2008) shows complex structure
perhaps indicating a merger of components responsible for the cold front. The caustics are cleanly determined.

{\it A2142:} A2142 is the prototype cold front cluster initially discovered by Markevitch et al. (2000). The weak lensing map by Umetsu et al. (2009) shows complex structure as for A2034.

{\it A2261:} A2261 contains a very unusual brightest cluster galaxy with a large central velocity dispersion and an extended flat core (Postman et al. 2012). The sampling of this cluster with the 300-fiber Hectospec instrument is restricted by bright stars in the field.  
Oddly, the hierarchical center of the cluster identified by the D99 procedure is $\sim 6^\prime$ south of the BCG and the BCG is
offset from the cluster mean by  about $\sim 400$ km s$^{-1}$ in the rest frame, as we reported in Coe et al. 2012. Serra et al. (2011) has a slightly different procedure for cutting the binary tree; this
algorithm yields the cluster center on the BCG. A2261 is a complex system and the D99 center
is coincident with a cluster substructure.
These two centers are relatively close to each other and 
the caustic mass profile is insensitive to the final center choice. Here we show the profile centered on the BCG.

The total sample of 19 clusters appears to include systems in a broad range of dynamical states. 
When examined in enough detail, every cluster of galaxies reveals some kind of complexity. Substructure enters the caustic and lensing estimates through its effect on the 
determination of the position of the caustics and its effect on the fitting of the NFW weak lensing profile. We note that the two clusters, A689 and A1758 with the largest ratios of caustic to weak lensing mass profiles at small radii are dominated by substructure and/or sampling issues. In spite of these issues,  we take all of the caustic and lensing profiles at face value to assess the relative measurements of cluster mass profiles in Section \ref{discussion}.

\section {MS0906: A Remarkable Example of Superposition Along the Line-of-Sight}
\label {ms0906}
Structures projected along the line-of-sight are a fundamental limitation on the accuracy of weak lensing profiles (e.g. Hoekstra 2001; Hoekstra 2003; White et al. 2002; de Putter \& White 2005; Hoekstra et al. 2011). Here we show a remarkable case where two massive clusters at different redshifts share nearly the same central position on the sky.  We demonstrate that this remarkable superposition
leads to very substantial contamination of the weak lensing mass profile published by Hoekstra (2007).
In fact, the weak lensing mass is approximately the sum of the masses of the two clusters nearly aligned along the line-of-sight.

MS0906+11 is an extended x-ray source discovered in the {\it Einstein} Slew 
Survey; Elvis et al. 1992). A750, a less luminous extended x-ray source, is centered only 5$^\prime$ (0.63 $h^{-1}$ Mpc) from the x-ray center of MS0906+11 (see Figure 3.39 of Maughan et al. 2008). The mean redshifts of the two clusters are similar: z = 0.1767 for MS0906+11 and z = 0.1640 for A750. On the basis of a sparse survey, Carlberg et al. (1996) wrote that MS0906 ``appears to be an indistinct binary in redshift space.'' 
 
\begin{figure}[htb]
\centerline{\includegraphics[width=7.0in]{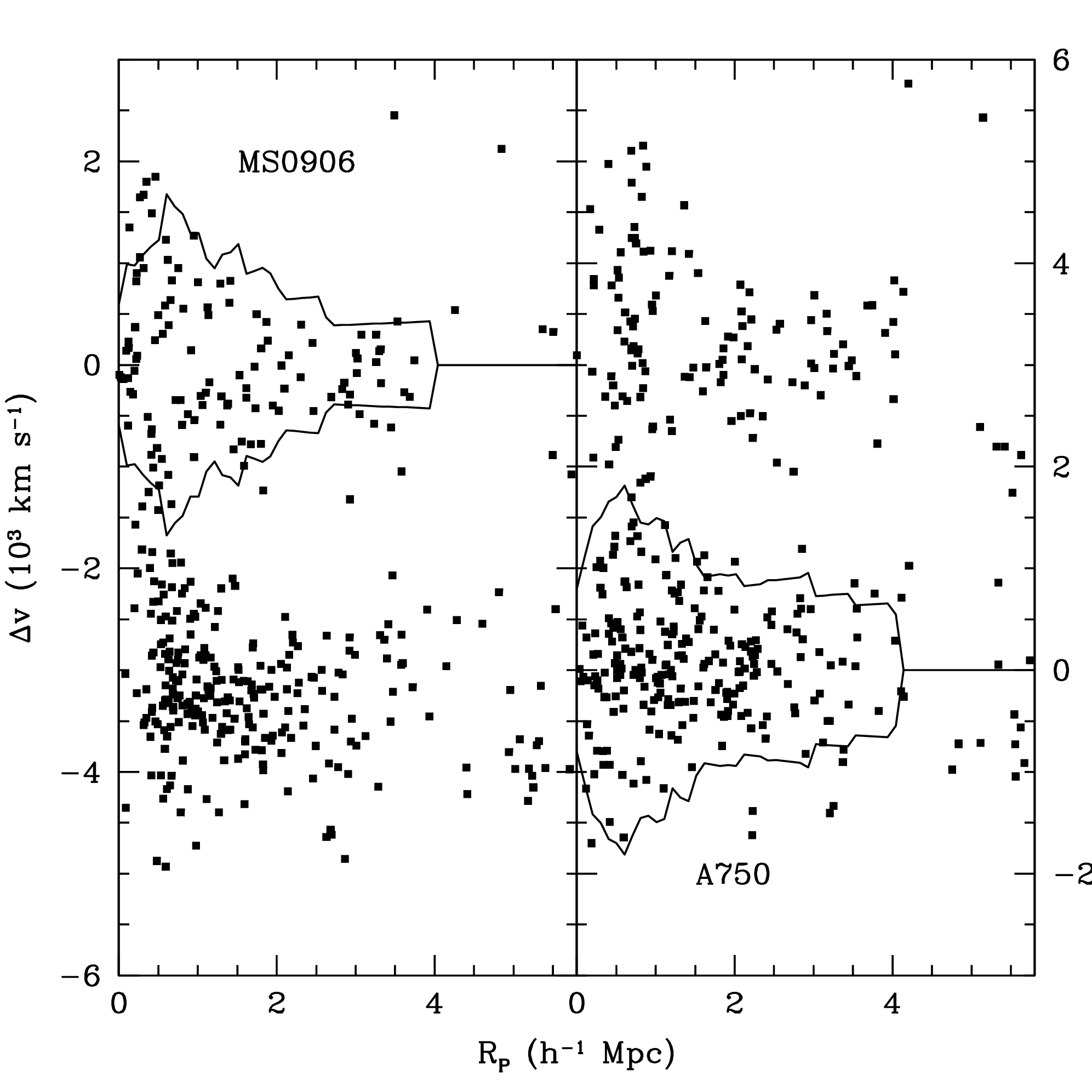}}
\vskip -1ex
\caption{Caustic diagrams for MS0906 and A750 superposition. On the left, MS0906 sets the zero-point for the relative line-of-sight velocities; on the right A750 sets the zero-point. 
\label{fig:caustic0906.ps}}
\end{figure}

Dense HeCS spectroscopy demonstrated cleanly that there are two distinct clusters along the line-of-sight (Rines et al. 2012). Figure \ref{fig:caustic0906.ps} shows the caustic diagrams for the two clusters; the two patterns are readily visible. In the left-hand panel MS0906+11 determines the zero point. In the right-hand panel, A750 determines the zero-point. In the rest frame the two clusters are separated by 3250 km s$^{-1}$.

There is some confusion in the literature about the identification of A750 and MS0906+11 on the sky.
Okabe et al. (2010) show maps of the surface mass density and galaxy red sequence luminosity density
for A750 (their Figure 30). Actually, in their panel B, the concentration marked C is MS0906+11,
not A750. Their NW1 component is A750. The cluster MS0906+11 has a greater central surface mass density (and presumably greater mass) than A750 (NW1); the 
red sequence luminosity density (galaxies on the red sequence with R$_{AB} < 22$) appears to be greater for A750. Because of the complexity of the system, Okabe et al. (2010) do not report a 
weak lensing mass for the apparently complex system thay call A750.

\begin{figure}[htb]
\centerline{\includegraphics[width=7.0in]{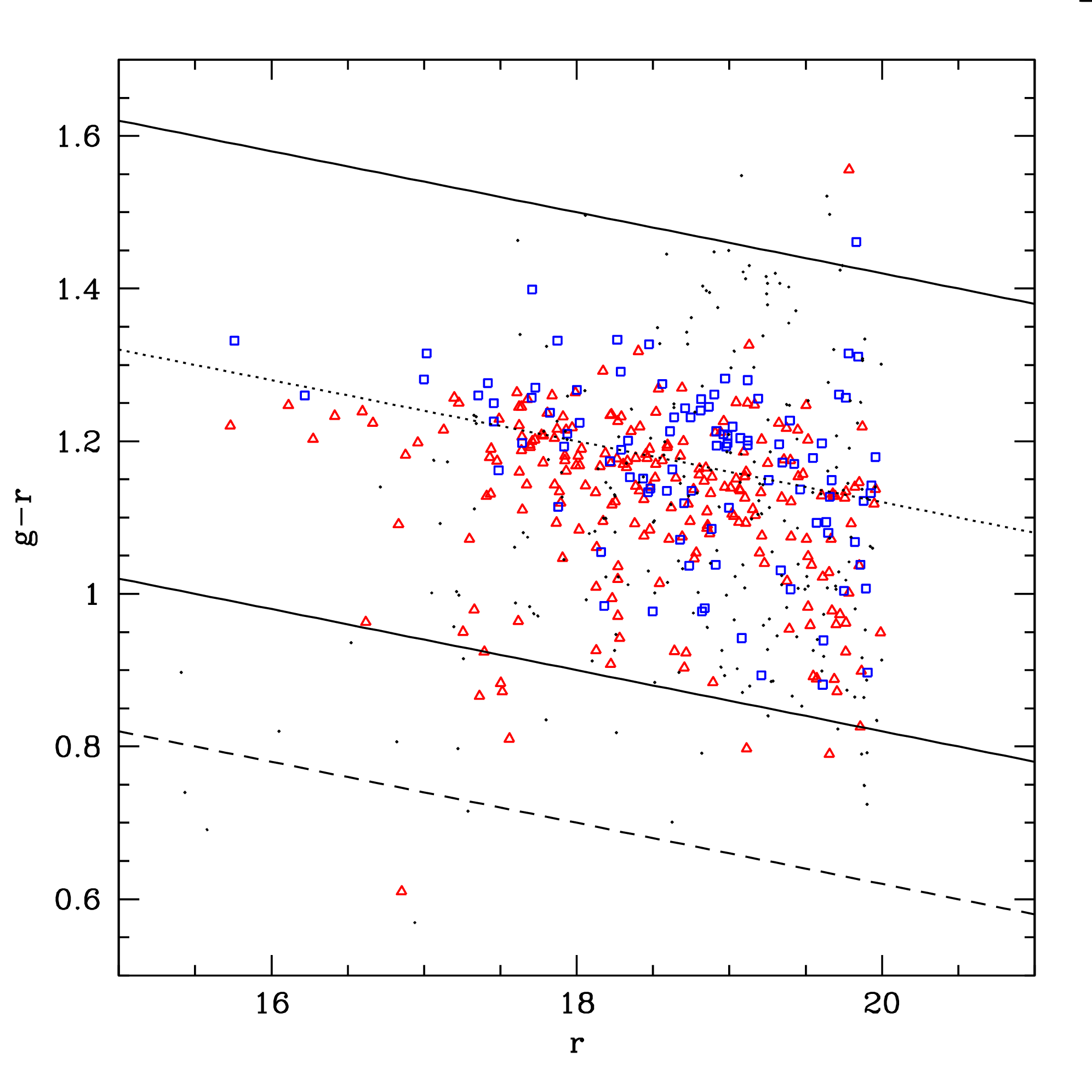}}
\vskip -2ex
\caption{Color-magnitude diagram for  MS0906/A750. {\it r} magnitudes and {\it g-r} color are from the SDSS. Solid lines indicate the main Hectospec target selection; the dashed lines indicate the limits for secondary targets. Red triangles indicate A750 members, blue squares are MS0906+11 members, and black dots are non-members.
\label{fig:cm0906.ps}}
\end{figure}

For the HeCS observations of the MS0906+11 field, we selected galaxies within 0.3 magnitudes of the red sequence at the mean redshift of MS0906+11 and with SDSS r = 16-21. Figure
\ref{fig:cm0906.ps} shows that the difference between the red sequences of MS0906+11 (blue open squares) and A750 (open red triangles) is subtle. We centered the  Hectospec pointings on the center of MS0906+11; thus 
any position bias (if any) in the spectroscopy  favors MS0906+11.

The HeCS data confirm the 
suggestion of Okabe et al. (2010); the number and total R-band luminosity of galaxies
in A750 exceeds those in MS0906+11. Within $R_{200}$ and with $M_r < -19$ there are 86 galaxies in A750 and only 41 in MS0906+11. The luminosity ratio is $L_{750}/L_{MS0906} \sim 1.55 \pm 0.29$, consistent with the mass ratio $M_{200, A750}/M_{200, MS0906}$ = 1.78$\pm$0.27.

\begin{figure}[htb]
\centerline{\includegraphics[width=7.0in]{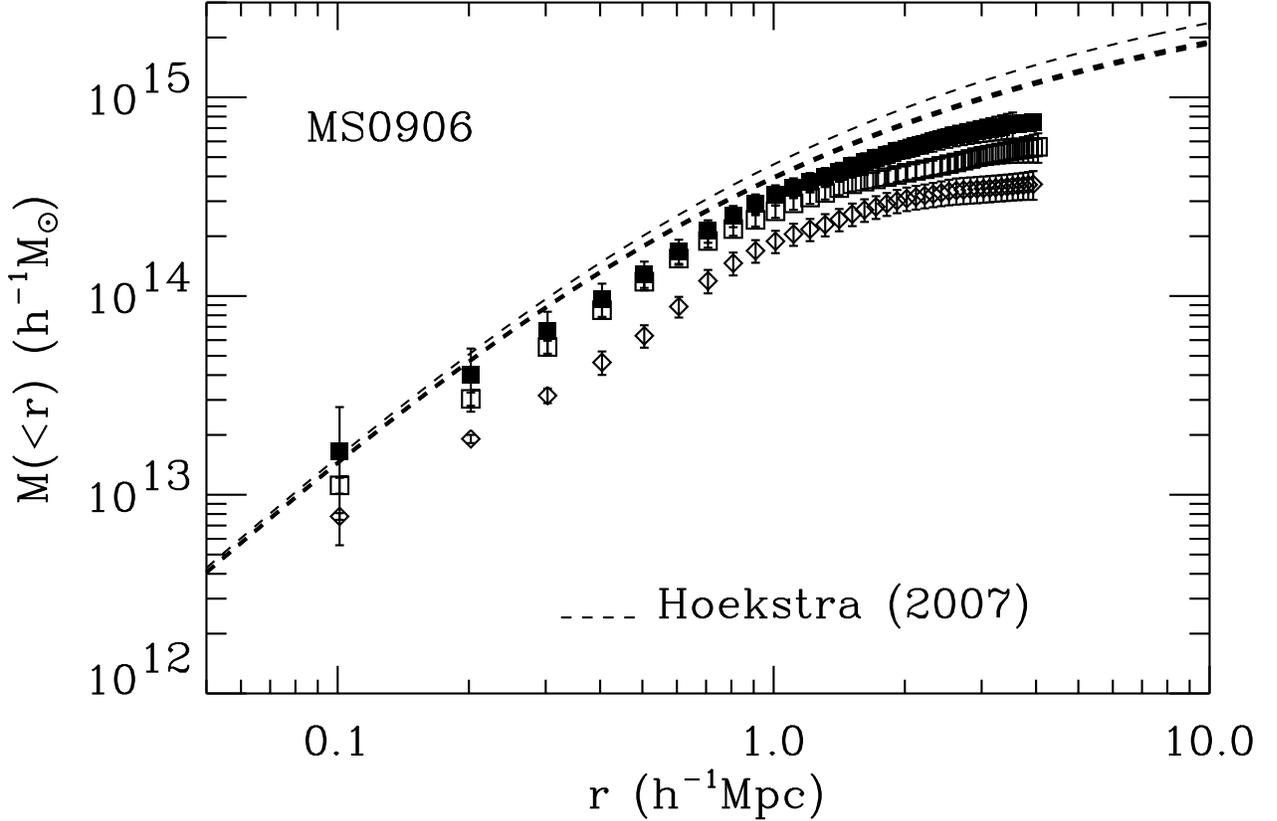}}
\vskip -2ex
\caption{Comparison of the weak lensing mass profile for MS0906 with 1$\sigma$ errors (heavy and light dashed curves respectively; Hoekstra et al. 2007) with the caustic mass profiles for MS0906 (open diamonds), A750 (open squares), and  the effective sum
caustic mass profiles for the two superposed clusters taking the 0.6$h^{-1}$ Mpc offset between the centers into account (solid squares). For A750, {$z = 0.164$},
{$M_{200}^{\rm caus}/10^{14}h^{-1}M_\odot = 2.61\pm 0.17$},
{$M_{200}^{\rm NFW}/10^{14}h^{-1}M_\odot = 2.64\pm 0.57$}, 
{$r_{200}^{\rm caus}/h^{-1}$Mpc $= 0.99\pm 0.03$}, 
{$r_{200}^{\rm NFW}/h^{-1}$Mpc $= 0.99\pm 0.07$}, 
{$c_{200}=r_{200}/r_{\rm rs} = 1.92\pm 0.24 $}, and
{$N_{\rm caus} = 225$}. 
\label{fig:mass0906.ps}}
\end{figure}

Figure \ref{fig:mass0906.ps} summarizes the published weak lensing mass determination for MS0906+11 and the
caustic masses derived from the HeCS data. For comparison with the lensing mass we
sum the caustic masses of A750 and MS0906+11, by taking into account the 0.6h$^{-1}$ Mpc projected
separation between the two cluster centers. The lensing mass is remarkably close to this sum of the dynamical masses computed from the caustic technique.

\section {Discussion}
\label {discussion}
Comparison of mass profiles derived from weak lensing and from the caustic technique highlights the 
relative strengths of the two techniques. Weak lensing measurements have small statistical errors at small radius. The major systematics at small radii (see Section \ref{weaklensing}) can be well controlled.
At large radius, where the lensing signal is much weaker, superposed structures can lead to bias in the determination of the mass profile. In contrast, for the caustic technique, the most severe bias occurs in the core where the assumption of a constant form factor (See Section \ref{caustics}) is poorest.
The caustic technique is insensitive to superposed structures throughout the radial range of the profile. 

At r$_{200}$, the typical error in a weak lensing mass profile is $\sim$ 20\% for a cluster at z = 0.3; the error increases to nearly $\sim$ 30\% at z = 0.1 (Hoekstra 2003; his Figure 7). Becker \& Kravtsov (2011) consider the systematic error introduced by fitting NFW profiles to weak lensing; depending upon the details of the weak lenisng analysis, the NFW fit underestimates the true mass by $\sim 5-10$\%. For the caustic mass profile the error at r$_{200}$ is $\lesssim 50$\%. For both techniques, the systematic error in this range appears to be relatively small.

\begin{figure}[htb]
\centerline{\includegraphics[width=7.0in]{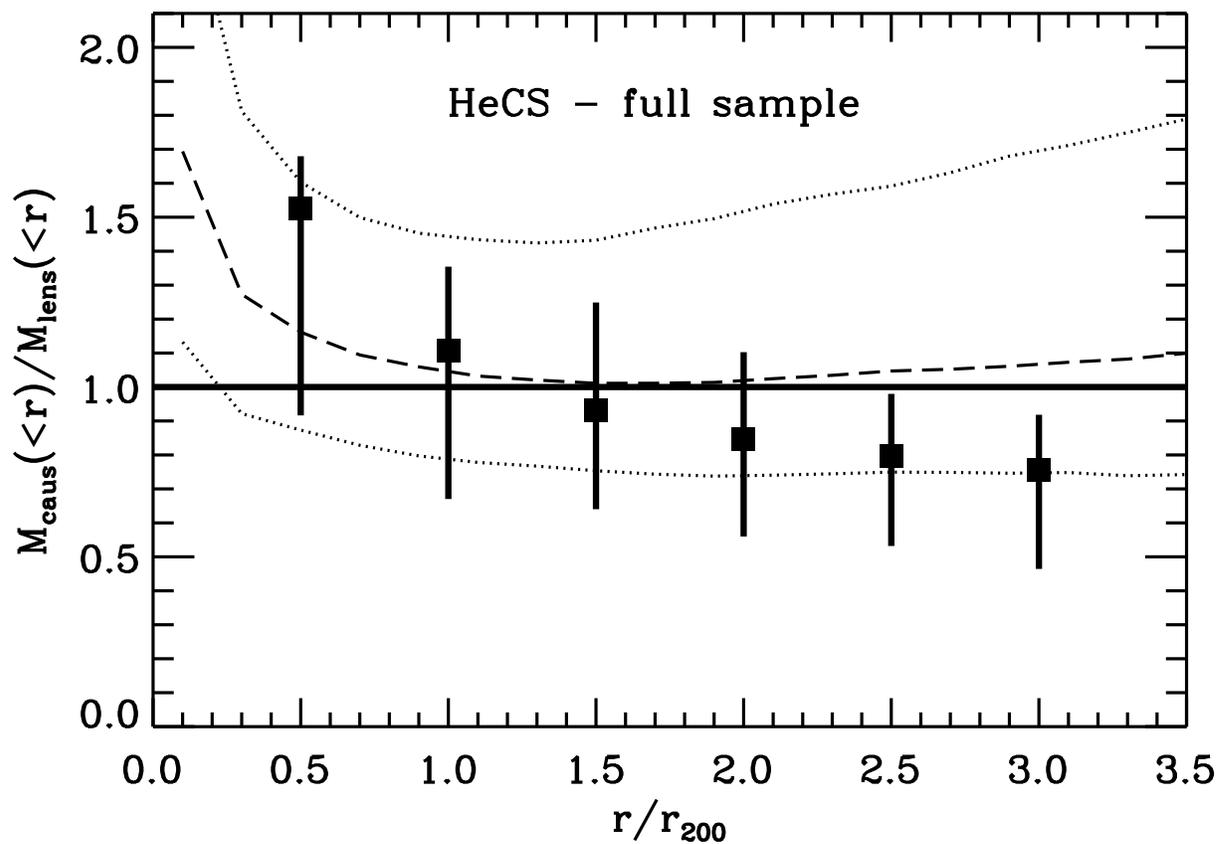}}
\vskip -2ex
\caption{Median (squares) ratio of caustic mass to weak lensing mass as a function of r/r$_{200}$.
The bars indicate the interquartile range. The dashed line shows the ratio between the caustic and true mass profiles derived from N-body simulations and the dotted lines show the 68\% confidence interval (we also plot these curves in Figures \ref{fig:caustics1.ps} --- \ref{fig:caustics7.ps}).
\label{fig:caustic-lens.ps}}
\end{figure}

Figure \ref {fig:caustic-lens.ps} shows the median behavior (and interquartile range) of the relative caustic and lensing mass profiles for our total sample of 19 clusters.  For comparison, the dashed line shows the median ratio between the caustic and true mass profiles of clusters extracted from an N-body simulation and the dotted lines show the 68\% confidence interval (we also plot these curves in Figures \ref{fig:caustics1.ps} --- \ref{fig:caustics7.ps}). Throughout the radial range we sample, the median ratio of the caustic and weak lensing profiles lies within the 68\% confidence intervals we derive from the 
N-body simulation.

Figure \ref {fig:caustic-lens.ps} shows that the weak lensing and caustic mass profiles agree stunningly near the virial radius (r$_{vir} \sim 1.3\rm{r}_{200}$). At radii less than r$_{vir}$ the
caustic mass profile exceeds the weak lensing profile. If the weak lensing profile is close to the true mass profile, the ratio between the caustic and weak lensing profiles behaves as we would expect based on the N-body simulations. In other words, if we chose an average form factor, F$_\beta$,  that is a function of radius (not a constant) based on the simulations, the weak lensing and caustic profiles would match 
to within $\sim 30$\% on scales less than r$_{vir}$. 

At radii greater than r$_{vir}$, the lensing profiles overestimate the mass profile relative to the
caustic estimate. This behavior suggests that superposed large-scale structure is indeed an increasing problem at large radii for weak lensing. At 3r$_{200}$, the comparison suggests that weak lensing overestimates the profile by $\sim 20-30$\% . 

At large radii, the contribution of large-scale structure to the weak lensing error budget is comparable with our estimate of the systematic error (see Hoekstra 2003). Some investigators have suggested that the error contributed by superposed large-scale structure is less important than claimed by Hoekstra (2003), but our results suggest that the impact of superposed large-scale structure is significant.  In another approach to evaluating the impact of large-scale structure on weak lensing measurements of mass profiles, Coe et al. (2012) remove superposed structures in their analysis of A2261; this removal decreases the viral mass by $\sim 7$\% suggesting that a substantial fraction of the $\sim$20\%
effect we see at larger radius might also result from superposed large-scale structure.

\section {Conclusion}

We compare cluster mass profiles derived with two fundamentally different methods. Weak lensing measures the projected surface mass density by analyzing small distortions of distant background galaxies (sources). Tha caustic technique is a kinematic technique based on the trumpet-like appearance of clusters in redshift space. Unlike a host of other  mass estimation techniques, these two approaches are independent of equilibrium assumptions. In principle they can both be applied over a large radial range.   

The nineteen clusters in our sample span the mass range $\sim10^{14}$ to 10$^{15}$h$^{-1}$M$_\odot$.
The median ratio of the caustic and weak lensing mass profile is within the 68\% confidence limits of the ratio between the true and caustic mass profiles derived from N-body simulations.  At radii $\lesssim r_{200}$, the caustic approach overestimated the mass, a behavior expected as a result of a constant form factor. Near the virial radius ($\sim 1.3 r_{200}$), the profiles agree to $\sim 30$\%. 

At large radius, weak lensing appears to systematically overestimate the mass profile by $\sim$ 20 - 30\%. We suggest that this apparent bias results from superposed large-scale structure within the lensing kernel. The caustic technique is insensitive to these structures. We demonstrate by examining the system MS0906+11 that the impact of superposed structures (including other clusters) can as large
as a factor of two. These results underscore the need for detailed simulations of potential biases 
poduced by large-scale structure superposed within the weak lensing kernel.
 
Spectroscopic data are rarely used in combination with weak lensing. Our analysis suggests that weak lensing mass profiles could be improved by using a redshfit survey to identify structures superposed along the line-of-sight and within the lensing kernel (see, for example, Coe et al. 2012). Furthermore the agreement of masses derived near the virial radius suggests that a combination of a dense cluster redshift survey with weak lensing estimates could be the basis for a more powerful method of assessing the mass distribution at  radii $\gtrsim r_{200}$ than either the caustic technique or weak lensing alone.

We thank Ian Dell'Antonio and Scott Kenyon for insightful discussions that helped us clarify issues in this paper. The Smithsonian Institution partially supports MJG's research.
KR was funded by a Cottrell College Science Award from the Research Corporation. ALS acknowledges a fellowship by the PRIN INAF09 ''Towards an Italian Network for Computational Cosmology''. AD and ALS acknowledge partial support from the INFN grant PD51 and the PRIN-MIUR-2008 grant ''Matter-antimatter asymmetry, dark matter and dark energy in the LHC Era''.

{\it Facilities:}\facility {MMT(Hectospec)}

\begin{deluxetable}{lcc}
\tablecolumns{3}
\tablewidth{0pc}
\tabletypesize{\footnotesize}
\tablenum{1}
\tablecaption{Weak Lensing Observations of HeCS and CIRS Clusters}
\tablehead{
\colhead{Cluster Name}&
\colhead{Telescope}&
\colhead{Reference}
}
\startdata 
A267&Subaru&Okabe et al.(2010)\\
 &CFHT&Hoekstra(2007)\\
A689&Subaru&Oguri et al.(2010)\\
A697&Subaru&Okabe et al.(2010)\\
MS0906&CFHT&Hoekstra(2007)\\
A963&Subaru&Okabe et al.(2010)\\
 &CFHT&Hoekstra(2007)\\
A1689&CFHT&Hoekstra(2007)\\
 &HST&Lemze et al.(2008)\\
A1750$^{*C}$&Subaru&Okabe and Umetsu(2008)\\
A1758$^*$&Subaru&Okabe and Umetsu(2008)\\
A1763&CFHT&Hoekstra(2007)\\
A1835&Subaru&Okabe et al.(2010)\\
A1914$^*$&Subaru&Okabe and Umetsu(2008)\\
A2009&Subaru&Okabe et al.(2010)\\
A2034$^*$&Subaru&Okabe and Umetsu(2008)\\
A2142$^{*C}$&Subaru&Umetsu et al.(2009)\\
A2219&Subaru&Okabe et al.(2010)\\
 &CFHT&Hoekstra(2007)\\
RXJ1720&Subaru&Okabe et al.(2010)\\
A2261&Subaru&Okabe et al.(2010)\\
A2631&Subaru&Okabe et al.(2010)\\
RXJ2129&Subaru&Okabe et al.(2010)\\
\enddata
\tablenotetext{*} {denotes a cluster {\it selected} for weak lensing observation as merging or cooling flow system.}
\tablenotetext {C} {denotes a cluster in the CIRS sample. All other systems are in the HeCS sample.}
\label{tbl:weaklens}
\end{deluxetable}

\clearpage
\begin{landscape}

\begin{deluxetable}{lccccccc}
\tablecolumns{8}
\tablewidth{0pc}
\tabletypesize{\footnotesize}
\tablenum{2}
\tablecaption{HeCS and CIRS Caustic Mass Estimates$^*$}
\tablehead{
\colhead{Cluster}&
\colhead{$z$}& 
\colhead{$M_{200}^{\rm caus}/10^{14}h^{-1}M_\odot$}&
\colhead{$M_{200}^{\rm NFW}/10^{14}h^{-1}M_\odot$}& 
\colhead{$r_{200}^{\rm caus}/h^{-1}$Mpc}& 
\colhead{$r_{200}^{\rm NFW}/h^{-1}$Mpc}& 
\colhead{$c_{200}=r_{200}/r_{\rm rs}$}& 
\colhead{$N_{\rm caus}$} \\
}
\startdata

A0267 &  0.229 & $ 4.92\pm 0.26$  & $ 7.5\pm 1.3$  & $ 1.19\pm 0.02$  & $ 1.37\pm 0.08$  & $ 1.80\pm 0.16$  & $   226$  \\
A0689 &  0.279 & $ 1.54\pm 0.05$  & $ 1.59\pm 0.07$  & $ 0.80\pm 0.01$  & $ 0.80\pm 0.01$  & $ 9.05\pm 0.35$  & $   163$  \\
A0697 &  0.281 & $ 4.4\pm 2.0$  & $ 7\pm 14$  & $ 1.13\pm 0.24$  & $ 1.33\pm 0.90$  & $
 1.1\pm 1.2$  & $   185$  \\
MS0906 &  0.177 & $ 1.47\pm 0.25$  & $ 0.95\pm 0.19$  & $ 0.81\pm 0.07$  & $ 0.70\pm 
0.05$  & $ 2.31\pm 0.26$  & $   101$  \\
A0963 &  0.204 & $ 4.00\pm 0.04$  & $ 4.35\pm 0.08$  & $ 1.12\pm 0.01$  & $ 1.15\pm 0.01$  & $ 3.69\pm 0.04$  & $   211$  \\
A1689 &  0.184 & $ 8.6\pm 1.9$  & $ 9.6\pm 5.8$  & $ 1.46\pm 0.12$  & $ 1.51\pm 0.30$ 
 & $ 7.1\pm 3.5$  & $   210$  \\
A1750$^C$ &  0.085 & $ 3.00\pm 0.09$  & $ 2.68\pm 0.31$  & $ 1.06\pm 0.01$  & $ 1.02 \pm 0.04$  & $ 2.69\pm 0.20$  & $   398$  \\
A1758 &  0.276 & $ 2.22\pm 0.77$  & $ 2.0\pm 1.2$  & $ 0.90\pm 0.13$  & $ 0.87\pm 0.17$ 
 & $ 4.9\pm 2.3$  & $   143$  \\
A1763 &  0.231 & $12.4\pm 1.4$  & $12.6\pm 3.8$  & $ 1.62\pm 0.08$  & $ 1.63\pm 0.16$ 
 & $ 5.7\pm 1.3$  & $   237$  \\
A1835 &  0.251 & $ 8.41\pm 0.53$  & $11.6\pm 1.3$  & $ 1.41\pm 0.03$  & $ 1.57\pm 0.06$  & $ 3.36\pm 0.23$  & $   219$  \\
A1914 &  0.166 & $ 4.75\pm 0.13$  & $ 4.62\pm 0.23$  & $ 1.20\pm 0.01$  & $ 1.19\pm 0
.02$  & $11.76\pm 0.63$  & $   255$  \\
A2009 &  0.152 & $ 3.49\pm 0.16$  & $ 3.16\pm 0.29$  & $ 1.09\pm 0.02$  & $ 1.05\pm 0
.03$  & $ 5.82\pm 0.44$  & $   195$  \\
A2034 &  0.113 & $ 5.00\pm 0.03$  & $ 5.60\pm 0.08$  & $ 1.25\pm 0.01$  & $ 1.29\pm 0.01$  & $ 3.00\pm 0.03$  & $   182$  \\
A2142$^C$ &  0.090 & $ 2.9\pm 1.0$  & $ 2.7\pm 3.2$  & $ 1.04\pm 0.18$  & $ 1.02\pm 0.41
$  & $ 2.5\pm 1.9$  & $   248$  \\
A2219 &  0.226 & $ 8.9\pm 1.9$  & $ 9.4\pm 5.2$  & $ 1.46\pm 0.14$  & $ 1.48\pm 0.27$  &
 $ 6.2\pm 2.8$  & $   461$  \\
RXJ1720 &  0.160 & $ 4.44\pm 0.24$  & $ 4.28\pm 0.43$  & $ 1.18\pm 0.03$  & $ 1.16\pm
 0.04$  & $ 9.6\pm 0.9$  & $   376$  \\
A2261 &  0.224 & $ 2.60\pm 0.73$  & $ 2.3\pm 2.4$  & $ 0.97\pm 0.12$  & $ 0.93\pm 0.32$ 
 & $ 2.2\pm 1.5$
  & $   228$  \\
A2631 &  0.277 & $ 3.77\pm 0.66$  & $ 3.8\pm 1.8$  & $ 1.07\pm 0.08$  & $ 1.07\pm 0.17
$  & $ 2.62\pm 0.79$  & $   173$  \\
RXJ2129 &  0.234 & $ 5.6\pm 1.0$  & $ 4.5\pm 1.7$  & $ 1.24\pm 0.11$  & $ 1.16\pm 0.15$ 
 & $ 9.6\pm 3.5$  & $   325$  \\

\enddata
\tablenotetext{*}{$M_{200}^{\rm caus}$ and $r_{200}^{\rm caus}$ from the caustic mass profile and the 
best-fit NFW parameters to the caustic mass profiles within $1 h^{-1}$~Mpc. The last column lists 
the number of galaxies within the caustics in the redshift diagram.}
\tablenotetext{C} {denotes a cluster in the CIRS sample. All other systems are in the HeCS sample.}
\label{tbl:causticmass}
\end{deluxetable}
\end{landscape}

\end{document}